\documentclass[preprint,eqsecnum,aps]{revtex4}
\usepackage{graphicx}
\usepackage{latexsym}
\usepackage{amsmath}
\usepackage{amssymb}

\def\barr{\begin{array}}
\def\earr{\end{array}}
\def\beq{\begin{equation}}
\def\eeq{\end{equation}}
\newcommand{\be}{\begin{equation}}
\newcommand{\ee}{\end{equation}}
\newcommand{\bea}{\begin{eqnarray}}
\newcommand{\eea}{\end{eqnarray}}
\newcommand{\bfig}{\begin{figure}}
\newcommand{\efig}{\end{figure}}

\newcommand{\dis}{\displaystyle}

\begin{document}

\title{\Large \bf Scalar Kaluza-Klein modes in a multiply warped braneworld}

\author{Ratna Koley $^{1, 2}$ \footnote{E-mail:ratna.koley@rediffmail.com},
Joydip Mitra $^1$ \footnote{E-mail:tpjm@iacs.res.in}
and
Soumitra SenGupta $^1$ \footnote{E-mail:tpssg@iacs.res.in}}
\affiliation{
$^1$Department of Theoretical Physics\\
Indian Association for the Cultivation of Science, \\
Kolkata  700 032, India \\
$^2$ Department of Physics, Bethune College, 181 Bidhan Sarani, Kolkata 700006, India}

\begin{abstract}
The Kaluza-Klein ( KK )modes of a massive scalar field on
a 3-brane embedded in six dimensional multiply warped spacetime are determined. 
Due to the presence of
warping along both the extra dimensions the KK mass spectrum splits
into two closely spaced branches which is a distinct feature of this model compared
to the five dimensional Randall-Sundrum model. This new cluster of
the KK mode spectrum is expected  to have interesting phenomenological 
implications for the upcoming collider experiments. Such a scenario
may also be extended for even larger number of orbifolded extra dimensions.
\end{abstract}

\maketitle

%%%%%%%%%%%%%%%%%%%%%%%%%%%%%%%%%%%%%%%%%%%%%%%%%%%%%%%%%%%%%%%%%%%%%%%%%%%%%%%%%%%%%%%%%%%%%%%%%%%%%%%%%
\section{Introduction}

The Randall-Sundrum (RS) model \cite{rs} was originally proposed to resolve the hierarchy
between the scale of weak and gravitational interactions, $m_W \sim 10^2~GeV$ and
$M_P \sim 10^{18}~ GeV$, respectively. The RS model is based on a truncated $AdS_5$ spacetime,
bounded by two 4D Minkowski walls, often called UV (Planck) and IR (TeV) branes. The curvature
in 5D induces a warped geometry on the brane which redshifts scales of
order $M_P$ at the UV brane to scales of order $m_W$ at the IR brane.
Much work has been carried out
on diverse aspects of such models in the last few years.
These include the attempted resolution of the hierarchy problem
{\cite{besan,rs}}, questions about the
localization of various types of fields on the brane {\cite{bajc}},
particle phenomenology in the context of braneworld {\cite{besan}}
and various other cosmological consequences {\cite{cosmo}}.
The actual existence of warped extra dimensions as well as a firm
foundational basis for these models still remain open issues.
However it has been widely recognised that one of the key signature for extra dimensions
can be obtained from the collider physics by exploring the contributions of various Kaluza-Klein ( KK ) modes 
in scattering amplitudes. Though the original RS model was formulated with only the gravity in the bulk,
various other models subsequently considered the implications of the bulk standard model fields and their
KK modes on the brane. The first step in this direction was to study the role of a scalar field
in the bulk.

In recent times several extensions of the RS model have been proposed with more than one extra 
dimension~\cite{6dmodel1,6dmodel2}. Most of these models consider several independent $S_{1}/Z_{2}$
orbifolds along with four dimensional Minkowski space-time.
Leblond {\em {et.al.}} showed a set of consistency conditions for braneworld scenarios with a
spatially periodic internal space in \cite{sumrule}, from which it is
derived that the necessity of a negative tension brane appears in five
dimensions only, it is not the prerequisite at higher dimensional
generalizations \cite{sumrule}. It is also apparent from the
multigravity scenario discussed in \cite{kogan} that the radion
stabilization problem and the presence of negative tension brane is an
artifact of five dimensional spacetime. The non-trivial curvature of
the internal space in case of two or more extra dimensions provides
the necessary bounce configuration of the warp factor without the need
of any negative tension brane \cite{6dmodel1}.

An interesting model has been proposed in an alternative scenario in~\cite{dcssg}
where the warped compact dimensions get further warped by a series of successive warping 
leading to \textit{multiply warped spacetime} with various p-branes sitting at the different
orbifold fixed points satisfying appropriate boundary conditions. 
In this scenario the lower dimensional branes including the standard model 3-brane exist at the 
intersection edges of the higher dimensional branes. The resulting geometry of the 
multiply warped D dimensional spacetime is given
by : $M^{1, D-1} \rightarrow \left\{ [M^{1,3} \times S^1/Z_2] \times S^1/Z_2 \right\} \times \cdots$, with $(D - 4)$ such 
warped directions. It has been argued that this multiply warped spacetime gives rise to interesting 
phenomenology and offers a possible explanation of the small mass splitting among the standard model fermions~\cite{dcssg, rsj1}. 
One of the interesting characteristics of such a model is the bulk coordinate dependence of the higher 
dimensional brane tensions. Such a coordinate dependent brane tension is shown
to be equivalent to a scalar field distribution on the higher dimensional brane
which constitute the bulk for the 3-branes located at the intersection edges of
these higher dimensional brane. This emerges naturally from the 
requirement of the orbifolded boundary conditions along the two compact directions \cite{dcssg}.
Such a scalar field distribution may have several interesting phenomenological 
significance  for the TeV brane physics \cite{dcssg}.

In this work we study the Kaluza-Klein modes of a 6-dimensional bulk scalar in the
multiply warped spacetime. A higher dimensional scalar, after compactification appears with its zero mode and various 
KK modes in the effective (3 + 1) dimensional theory. 
We undertake to study both massless and massive modes for a bulk scalar in the present work.

The present article is organized as follows. In section II, we briefly discuss about the multiply warped six dimensional brane model. 
Then in section III we focus on the bulk scalar and  find out it's equation of motion and the Kaluza Klein modes. We conclude with
discussions in section IV.

\section{model}

The spacetime that we are interested in is a doubly warped and
compactified six-dimensional one with a $Z_2$ orbifolding in each of
the extra dimensions \cite{dcssg}. We use the following  notations for our discussion. The non-compact directions 
are denoted by $x^{\mu} (\mu = 0...3)$  and the orbifolded compact directions are
represented by the angular coordinates $y$ and $z$ respectively with
$R_y$ and $r_z$ as respective moduli.  The corresponding metric for
the six dimensional spacetime is given by

\be
ds^2_{(6)} = b^2(z) \left[a^2(y) \eta_{\mu \nu} dx^{\mu} dx^{\nu} + R^2_y dy^2 \right] + r^2_z dz^2
\label{metric}
\ee

where $\eta_{\mu \nu} = diag(-1, 1, 1, 1)$. The warp factors due to the extra
dimensions $y$ and $z$ are given by the functions $a(y)$ and $b(z)$ respectively.
Since orbifolding requires, in general, a localized concentration of energy, the
four branes are considered to be located at the orbifold fixed points, namely $y = 0,\pi$ and $z = 0, \pi$. The 
braneworld model is thus constructed from the
following action (\ref{Action}) which consists of a negative bulk cosmological constant and
coordinate dependent brane tensions. The total bulk-brane action is thus given by,

\be
\barr{rcl}
S & = & \dis S_6 + S_5 + S_4 \\[1ex]
S_6 & = & \dis \int {d^4 x} \, {d y} \, {d z} \,
          \sqrt{-g_6} \; \left(R_6 - \Lambda \right)
\\[2ex]
S_5 & = & \dis \int {d^4 x} \, {d y} \, {d z} \,
           \left[ V_1 \, \delta(y) + V_2 \, \delta( y - \pi) \right]
\\[1.5ex]
    & + & \dis \int {d^4 x} \, {d y} \, {d z} \,
           \left[ V_3 \, \delta(z) + V_4 \, \delta(z - \pi) \right]
\\[2ex]
S_4 & = & \dis \int d^4 xdydz \sqrt{-g_{vis}}[{\cal L} - \hat V] \ .
\earr
    \label{Action}
\ee

where $\Lambda$ is the bulk cosmological constant which is necessarily negative. In general, the brane potential 
terms (brane tensions) are $V_{1,2} = V_{1, 2}(z)$ whereas $V_{3, 4} = V_{3, 4}(y)$. The contributions due to possible 3-branes 
located at $(y, z) = (0,0), (0, \pi), (\pi, 0), (\pi, \pi)$ are
indicated by the term $S_4$. The solution of the Einstein equation for the metric
(\ref{metric}) in this action (\ref{Action}) leads to the warp factors in the following form :

\bea
a(y) = e^{-c \vert y \vert} ~~~~~~~~ c = \frac{R_y k}{r_z \cosh(k \pi)} \\ \nonumber
b(z) = \frac{\cosh(kz)}{\cosh(k\pi)} ~~~~~~~~~~ k = r_z \sqrt{\frac{- \Lambda}{M^4}}
\label{soln:interm}
\eea

Note that the solutions are $Z_2$ symmetric about $y$ and $z$
directions. One can obtain the brane tensions by considering the boundary
terms. The brane tensions at the two boundaries $y = 0 $ and $y= \pi$ are given by

\be
V_1(z) = - V_2(z) = 8M^2 \sqrt{\frac{-\Lambda}{10} } \,
{\rm sech}(k \, z) \ .
\ee

In other words, the two 4-branes sitting at $y = 0$ and $y = \pi$ have
$z$-dependent tensions. Similarly, the boundary condition for the
infinitesimal interval across $ z = 0 $ and $z = \pi$ leads to
\bea
V_3(y) & = & 0 \\ \nonumber
V_4(y) & = & -\frac{8 \, M^4 \, k}{r_z} \, \tanh(k\pi)
\eea

Note that, in this case the brane tensions are constants unlike the previous case, but quite similar to the case for the 
original RS model. The fact of $g_{yy}$ being a non-trivial function of $y$, however, made it mandatory that the two 
hypersurfaces accounting for the $y$-orbifolding must have a $z$-dependent energy density. This is, in fact, the most interesting part of
the model that we are currently interested in. If there exists no other
brane with a natural energy scale lower than ours, we must identify the SM brane with the one at $y= \pi, z = 0$.

In this model the solution of hierarchy problem
({\em i.e.} the mass rescaling due to warping) demands that unless there is a large hierarchy between the 
moduli $r_z$ and $R_y$, either of $c$ and $k$ must be small implying large warping in one direction and small in the other.
This particular feature of this model is revealed from the relation $c = \frac{R_y k}{r_z \cosh(k \pi)}$, which
implies that for $R_y \sim r_z$ a large hierarchy in the
$y$-direction (a situation very close in spirit with RS)
necessarily leads to a a relatively small $k \sim {\cal O}(1)$ and
hence little warping in the $z$-direction.

In summary we are dealing with a braneworld which is doubly warped and the warping is 
large  along one direction and small in the other. We now address the nature of the Kaluza-Klein modes of a six dimensional bulk
scalar field in such a braneworld.

\section{scalar modes}

We start with a six dimensional free massive scalar, $\Phi(x_{\mu}, y, z)$ of mass $m_{\phi}$. The action for the scalar is

\be
S_\Phi = \frac{1}{2} \int d^4 x \int^{\pi}_{-\pi} dy \int^{\pi}_{-\pi} dz \left[-g^{MN} \partial_{M} \Phi \partial_{N} \Phi - m^{2}_{\phi} \Phi^{2} \right]
\label{scalar:action}
\ee

where $g_{MN}$ is the six dimensional metric and
$m_{\phi}$ is of the order of Planck mass. We perform a Kaluza Klein decomposition of the
scalar as a sum over the modes.

\be
\Phi(x^{\mu}, y, z) = \sum_{np}\frac{ \phi_{np} (x^{\mu}) \xi_{n} (y) \chi_{p}(z)}{\sqrt{R_y r_z}}
\ee

%If $\xi_{n} (y)$ and  $\chi_{p}(z)$ are chosen to satisfy the following equations

%\beq\label{ymode}
 %\int \left(- \frac{\xi_{n_1}(y) \partial^{5} (a^{4}(y) \partial_{5}
 %\xi_{n_2}(y))}{R_y^2} + m_{n}^2 a^{4}(y) \xi_{n_1}(y) \xi_{n_2}(y) \right) dy = \int
 %m^{2}_{np} a^{2}(y) \xi_{n_1}(y) \xi_{n_2}(y) dy  
%\eeq
%\beq
% \int \left( - \frac{\chi_{p_1}(z) \partial^{6} (b^{5}(z) \partial_{6}
% \chi_{p_2}(z))}{r_z^2} + m_{\phi}^2 b^{5}(z) \chi_{p_1}(z) \chi_{p_2}(z) \right) dz =
% \int m^{2}_{n} b^{3}(z) \chi_{p_1}(z) \chi_{p_2}(z) dz 
%\label{zmode}
%\eeq
Substituting this in the six dimensional action one arrives at a four dimensional action for free massive scalar field with 
canonical kinetic term in the scalar field action provided the following normalization conditions are satisfied,
\bea
\int a^2(y) \xi_{n_{1}}(y) \xi_{n_{2}}(y) dy = \delta_{n_{1} n_{2}} \\
\int b^3(z) \chi_{p_{1}}(z) \chi_{p_{2}}(z) dz = \delta_{p_{1} p{2}}
\eea

Under this compactification the four dimensional effective action of the scalar field becomes

\be
S^{eff}_{4D}(\phi) = - \frac{1}{2} \sum_{np} \int \left( \partial^{\mu} \phi_{np}  \partial_{\mu} \phi_{np} + m^2_{np} \phi^{2}_{np} \right) d^4 x
\ee
As in usual Kaluza-Klein (KK) compactifications, the bulk field $\Phi(x^{\mu}, y, z)$ manifests itself to a four-dimensional 
observer as an infinite ``tower'' of scalar modes $\phi_{np} (x)$ with mass $m_{np}$. Note that, the KK mass term carries 
two indices because of the  two compact warped extra dimensions. This in turn means that the usual five dimensional massive tower  
here further splits into 
further sub-tower because of the additional warped dimension. These extra modes naturally will have their contributions in the 
particle collider experiments and are expected to produce enhanced signature for the extra dimensions. 

From the action we find the following eigenvalue equations for the 
$y$  and $z$ part of the the scalar field. The equation for the  `$y$' dependence  of the scalar is given by

\be
 \frac{1}{R_y^2} \dfrac{d}{dy} \left(a^4 \dfrac{d  \xi_{n}(y)}{d y}\right)
- m^2_{p} a^2 \xi_{n}(y) = - m^2_{np} a^4 \xi_{n}(y)
\label{yeq}
\ee

while the equation corresponding to the `$z$' dependence turns out to be

\be
\frac{1}{r_z^2} \dfrac{d}{dz} \left(b^5 \dfrac{d  \chi_{p}(z)}{d z} \right) - m^2_{\phi} b^5 \chi_{p}(z) = - m^2_{p} b^3 \chi_{p}(z)
\label{zeq}
\ee

The mass tower due to the extra dimension along $z$ direction is given by
$m_p$ whereas the four dimensional KK mass tower is represented by $m_{np}$.
It is interesting to note from the above equations that like 5-dimensional case, the bulk scalar mass appears in the
equation for $ \chi_{p}(z)$ to determine the parameter $m_p$ whereas   $m_p$  
enters into the equation for  $\xi_{n}(y)$ and determines $m^2_{np}$ i.e. various KK mode masses in terms of the
two KK numbers $p$ and $n$.
Therefore to obtain the KK mass spectrum we first determine the tower denoted by $m_p$ and then using it as
the input to the Eq. (\ref{yeq}) we finally achieve to obtain the desired mass spectrum in the 3-dimensional 
visible brane. Considering the allowed domain for the values of $k$ we consider
the following approximated form
of the warp factor $ b(z) \sim e^{-k (\pi - z)} = e^{- k \tilde z}$. Now redefining the variable as
$z_{p} = \frac{m_p}{k'} e^{k \tilde z}$  where $k' = \frac{k}{r_{z}}$ and the function as
$\tilde \chi_p (z) = e^{- \frac{5}{2} k \tilde z} \chi_{p} (z)$, the Eq. (\ref{zeq}) can be recast as

\be
{z_p}^2 \dfrac{d^2 \tilde \chi_{p}}{d z_{p}^2} + z_p \dfrac{d \tilde \chi_{p}}{d z_p} + ({z_p}^2 - \nu_{\phi}^2) \tilde \chi_{p} = 0
\label{zpeq}
\ee

where $\nu_{\phi}^2 = \left( \frac{m_{\phi}^2}{k'^2} + \frac{25}{4} \right)$ , $m_{\phi}$ being
the bulk mass of the scalar field. The solutions of the above equation are the Bessel functions of order
$\nu_{\phi}$

\be \chi_{p} (z) = N_p e^{\frac{5}{2} k \tilde z}
\left[J_{\nu_{\phi}}(\frac{m_p}{k'} e^{k \tilde z})  + b_{p}
Y_{\nu_{\phi}}(\frac{m_p}{k'} e^{k \tilde z}) \right] \label{zpart}
\ee

where $N_{p}$ is the normalization constant and $b_{p}$ is an
arbitrary constant.
%Apply boundary conditions .... Bessel function equation ..... mass spectrum

Following the condition that the left hand side of the Eq.
(\ref{zpeq}) is self-adjoint,  the first order derivative of
$\chi_{p}(z)$ must be continuous at the orbifold fixed points $z= 0$
and $z = \pi$.  This leads us to the spectrum for $m_p$. From the condition of self-adjointness, we approximately get a condition 
\be \frac{5}{2} J_{\nu_{\phi}} (x_{p \nu_{\phi}}) + x_{p \nu_{\phi}}~ J_{\nu_{\phi}}' (x_{p \nu_{\phi}}) = 0
\label{6dmode} \ee
where $ x_{p \nu_{\phi}}=\frac{m_p}{k'}e^{k \pi}$.
\\ After obtaining the $z$ dependent part of the scalar KK modes and the corresponding
spectrum we solve the $y$ dependent part of the modes and finally
arrive at the full spectrum of the KK modes from the point of view of a 
3-brane observer sitting at the visible brane.

The solution of Eq.(\ref{yeq}) yields $\xi_{n}
(y)$ as Bessel function of order $ \nu_p = \sqrt{4 +
\frac{{m_{p}}^{2}}{k'^{2}}} \label{nu} $ multiplied by a growing
exponential factor as,

\be \xi_{n} (y) = N_{n} e^{2 c \vert y \vert} \left[ J_{\nu_p}
\left(\frac{m_{np}}{k'} e^{c \vert y \vert}\right) + b_{n} Y_{\nu_p}
\left( \frac{m_{np}}{k'} e^{c \vert y \vert} \right) \right]
\label{ypart} \ee

Here $N_{n}$ is the normalization constant and $b_{n}$ is an arbitrary constant.

Once again following the condition that the left hand side of the Eq. (\ref{yeq}) is self-adjoint, the first order 
derivative of
$\xi_{n}(y)$ must be continuous at the orbifold fixed points. This gives us the
equations from which we obtain the mass spectrum, $m_{np}$.
Applying the above condition in the Eq. (\ref{ypart}) at the location of the visible brane at $y = \pi$ and using $e^{c \pi} >> 1$
we obtain the following transcendental equation:

\be 2 J_{\nu_p} (x_{n p}) + x_{n p}~ J_{\nu_p}' (x_{n p}) = 0
\label{5dmode} \ee

where,  $x_{n p} = m_{n p} e^{c \pi} / k'$. The mass spectrum
$m_{np}$ thus is  obtained from the solution of the above
transcendental equation.

In the five dimensional case the order of the Bessel function
depends on the mass of the scalar field. Therefore for a given bulk
scalar the KK modes are given by the wave functions (Bessel
functions) which are of a \emph{fixed order}. However in the six
dimensional case we can have several orders of the Bessel function
representing the KK modes because the order $\nu_p$ can take up
different values for different values $p$ \emph{i.e.} for different
values of $m_p$. Therefore for each mass splitting due to $z$
direction we will obtain spectrum of KK modes. Hence we obtain extra
splitting in the spectrum over the usual 5-dimensional scenario. In the following table
we show explicitly the KK modes for two given values of bulk scalar mass,
\begin{table}[h]
\begin{tabular}{|c|c|c|c|c|c|c|c|c|}
   \hline 
%& \multicolumn{1}{|c|}{n} & \hline
%& \multicolumn{4}{|c|}{p ------>} & \multicolumn{4}{|c|}{p ---->} & \hline
 &  \multicolumn{4}{|c|}{$m_{\phi} = 0.01 k'$}  & \multicolumn{4}{|c|}{$m_{\phi} = k'$} \\ 
\cline{1-9}
  $\nu_p$ & $\nu_1$ & $\nu_2$ & $\nu_3$ & $\nu_4$ & $\nu_1$ & $\nu_2$ & $\nu_3$ & $\nu_4$ \\ \hline
  $m_{1p} $& 5.04 & 6.61 & 8.26 & 9.99 & 5.11 & 6.69 & 8.34 & 10.05 \\            \hline
  $m_{2p}  $ & 8.44 & 10.26 & 12.14 & 13.98 & 8.53 & 10.36 & 12.23 & 14.15 \\          \hline
  $m_{3p} $ & 11.69 & 13.62 & 15.61 & 17.56 & 11.79 & 13.72 & 15.71 & 17.73 \\           \hline
  $m_{4p}$   & 14.89 & 16.89 & 18.95 & 20.97 & 14.99 & 17.00 & 19.06 & 21.15 \\          \hline
  \end{tabular}
  \caption{The KK mode masses $m_{np}$ (in TeV) for different values of $n$ and $p$. }
  \end{table}
 
Once again, like the 5-dimensional case, the scalar KK mode masses are suppressed by the warp factor. Taking
$m_{\phi}$ of the order of Planck scale we find that the light KK modes
have masses in the range of TeV. The exponential supression can be
understood from the equation of $\xi_{n}(y)$ which shows that the
modes are peaked near the visible brane at $\{y = \pi, z = 0\}$ \cite{rsj1}.
However in this case a much larger number  of KK modes of mass $\sim$ Tev appear in comparison to its
5-dimensional counterpart  and are expected to produce additional contributions to various processes involving KK modes 
of the bulk fields in the forthcoming collider experiments
at TeV scale.  
\section{conclusion}
In this work we have calculated various Kaluza-Klein mode masses of a bulk scalar field in a braneworld model with two warped 
extra dimensions.
Such bulk scalars are useful candidates for moduli stabilization of warped braneworld models \cite{gw}.
We have shown an increase in the number of scalar KK modes within a energy range of few Tev due to
\begin{itemize}
 \item 
the presence of
additional extra dimension over the usual five dimensional RS model

and 

\item
the specific characteristics of such
a model where a large warping in one direction is accompanied by a small warping in the other direction.
\end{itemize}
 
Such enhancement of the number of KK modes will take place for the other standard model
fields also if they are allowed to propagate in the bulk. Extending this work for 
even more larger number of warped dimensions \cite{dcssg} where two clusters of 3-branes with energy scale close
to TeV scale and Planck scale are obtained, it is easy to show a very large proliferation of the number of KK modes for each bulk field,
with masses close to Tev scale.
Such an increase in the number of KK modes is expected to modify the decay as well as the scattering amplitudes of
different processes in a Tev-scale collider. Their signature therefore may play a crucial role in determining
the number of warped directions in our search for extra dimensions in the collider experiments.

\acknowledgments{JM acknowledges Council for Scientific and
Industrial Research, Govt. of India for providing financial support.
RK thanks the Department of Science and Technology (DST), Government
of India, for financial support through the project no.
SR/FTP/PS-68/2007. RK also thanks IACS, Kolkata where the major part of the
project has been done.}


\newpage

\begin{thebibliography}{99}

\bibitem{rs} L. Randall and R. Sundrum, Phys. Rev. Lett. {\bf 83}, 3370 (1999)


\bibitem{besan} M. Besancon, {\em Experimental introduction to
extra dimensions}, hep-th/0106165 ; \newline  Y. A. Kubyshin, {\em Models with
extra dimensions and their phenomenology} hep-ph/0111027
and references therein.

\bibitem{bajc} B. Bajc and G. Gabadadze, Phys. Lett. {\bf B 474}, 282
  (2000);  \newline S. L. Dubovsky, V. A. Rubakov and P. G. Tinyakov,
  Phys. Rev. {\bf D 62}, 105011 (2000)



\bibitem{cosmo} D. Langlois, gr-qc/0102007 (Proceedings of the
9th Marcel Grossmann meeting, July 2000 (Rome)); \newline  gr-qc/0205004 (Proceedings
of Journees Relativistes, Dublin 2001) and references therein
; \newline  E. Flannagan, S. H. Henry Tye and I. Wasserman,
Phys. Rev. {\bf D 62}, 024011 (2000);  \newline H. Stoica, S. H. Henry Tye
and I. Wasserman, Phys. Letts. {\bf B 482}, 205 (2000);  \newline J. S. Alcaniz
, astro-ph/0202492


\bibitem{6dmodel1} A.~G.~Cohen and D.~B.~Kaplan,
  %``Solving the hierarchy problem with noncompact extra dimensions,''
  Phys.\ Lett.\  B {\bf 470}, 52 (1999);
  R.~Gregory,
  %``Nonsingular global string compactifications,''
  Phys.\ Rev.\ Lett.\  {\bf 84}, 2564 (2000);
  S.~M.~Carroll, S.~Hellerman and M.~Trodden,
  %``BPS domain wall junctions in infinitely large extra dimensions,''
  Phys.\ Rev.\  D {\bf 62}, 044049 (2000);
  N.~Arkani-Hamed, L.~J.~Hall, D.~Tucker-Smith and N.~Weiner,
  %``Solving the hierarchy problem with exponentially large dimensions,''
  Phys.\ Rev.\  D {\bf 62}, 105002 (2000);
  Z.~Chacko and A.~E.~Nelson,
  %``A solution to the hierarchy problem with an infinitely large extra
  %dimension and moduli stabilization,''
  Phys.\ Rev.\  D {\bf 62}, 085006 (2000);
  T.~Gherghetta and M.~E.~Shaposhnikov,
  %``Localizing gravity on a string-like defect in six dimensions,''
  Phys.\ Rev.\ Lett.\  {\bf 85}, 240 (2000).



\bibitem{6dmodel2} M.~Giovannini, H.~Meyer and M.~E.~Shaposhnikov,
  %``Warped compactification on Abelian vortex in six dimensions,''
  Nucl.\ Phys.\  B {\bf 619}, 615 (2001); P.~Kanti, R.~Madden and K.~A.~Olive,
  %``A 6-D brane world model,''
  Phys.\ Rev.\  D {\bf 64}, 044021 (2001); C.~P.~Burgess, J.~M.~Cline, N.~R.~Constable and 
H.~Firouzjahi,
  %``Dynamical stability of six-dimensional warped brane-worlds,''
  JHEP {\bf 0201}, 014 (2002); M.~Giovannini,
  %``Gauge field localization on Abelian vortices in six dimensions,''
  Phys.\ Rev.\  D {\bf 66}, 044016 (2002); M.~Giovannini,
  %``Scalar normal modes of higher dimensional gravitating kinks,''
  Class.\ Quant.\ Grav.\  {\bf 20}, 1063 (2003)

\bibitem{sumrule} F. Leblond, R. C. Myers and D. J. Winters, JHEP
  {\bf{0107}} 031 (2001)

\bibitem{kogan} I. I. Kogan, S. Mouslopoulos, A. Papazoglou and
  G. G. Ross, Phys. Rev. {\bf D 64}, 124014 (2001)


\bibitem{dcssg}  D.~Choudhury and S.~SenGupta,
  %``Living on the edge in a spacetime with multiple warping,''
  Phys.\ Rev.\  D {\bf 76}, 064030 (2007)

\bibitem{rsj1} R.~Koley, J.~Mitra and S.~SenGupta,
  %``Chiral fermions in a spacetime with multiple warping,''
  Phys.\ Rev.\  D {\bf 78}, 045005 (2008)

\bibitem{gw} W.~D.~Goldberger and M.~B.~Wise,
  %``Bulk fields in the Randall-Sundrum compactification scenario,''
  Phys.\ Rev.\  D {\bf 60}, 107505 (1999).

%\bibitem{multiloc} I.~I.~Kogan, S.~Mouslopoulos, A.~Papazoglou and G.~G.~Ross,
  %``Multi-localization in multi-brane worlds,''
  Nucl.\ Phys.\  B {\bf 615}, 191 (2001)

\end{thebibliography}
\end{document}